\documentclass[12pt]{article}

\usepackage{amsmath}
\usepackage{amssymb}
\usepackage{latexsym}
\usepackage{graphicx}
\usepackage{psfrag,fancyhdr,epsfig}

\def\beq{\begin{equation}}
\def\eeq{\end{equation}}
\def\bea{\begin{eqnarray}}
\def\eea{\end{eqnarray}}

\begin{document}
\begin{titlepage}

\vspace*{1cm}

\begin{center}
{\bf {\Large {Scalar Emission in the Bulk in a Rotating \\[2mm] Black Hole
Background}}}

\bigskip \bigskip \medskip

{\bf S. Creek}$^1$, {\bf O. Efthimiou}$^2$, {\bf P. Kanti}$^{1,2}$
and {\bf K. Tamvakis}$^2$

\bigskip
$^1$ {\it Department of Mathematical Sciences, University of Durham,\\
Science Site, South Road, Durham DH1 3LE, United Kingdom}

$^2$ {\it Division of Theoretical Physics, Department of Physics,\\
University of Ioannina, Ioannina GR-45110, Greece}

\bigskip \medskip
{ \bf{Abstract}}
\end{center}

We study the emission of scalars into the bulk from a higher-dimensional
rotating  black hole. We obtain an analytic solution to the field equation
by employing matching techniques on expressions valid in the near-horizon
and far-field regimes. Both analytic and numerical results for the absorption
probability, in the low-energy and low-angular momentum limit, are derived
and found to be in excellent agreement. We also compute
the energy emission rate, and show that the brane-to-bulk ratio of the
energy emission rates for scalar fields remains always larger than unity in the
aforementioned regime.

\end{titlepage}

\setcounter{page}{1} \noindent

\section{Introduction}
Theories of extra spatial dimensions \cite{ADD, RS} offer
hope for the resolution of the hierarchy problem arising between
the scale of gravity and that of other fundamental interactions.
Within this framework gravity, and possibly scalar fields, propagate
in the full $(4+n)$-dimensional continuum (bulk), while ordinary
matter is trapped on a four-dimensional hypersurface (brane).
The resolution of the hierarchy problem arises because the fundamental
scale of higher-dimensional gravity can be close to the other particle
physics scales, while the traditional scale of gravity (Planck mass)
is related to it in terms of the volume and number of extra dimensions.
In this context, \textit{trans-planckian} collisions can give
rise to potentially observable black hole creation \cite{creation}.
These higher dimensional black holes \cite{MP} could be produced
either in ground-based colliders or in cosmic ray interactions and
detected through the emitted Hawking radiation \cite{hawking}.
There is a considerable amount of recent literature on the subject
(see \cite{reviews}) including both numerical and analytic studies
of the Hawking radiation in the Schwarzschild phase
\cite{KMR1, HK1, Barrau, Jung, BGK, Naylor, Park, Cardoso, CEKT1, Dai}
as well as the spin-down phase \cite{FS-rot, IOP, rot-bulk, HK2, IOP2,
Jung-super, Jung-rot, DHKW, CKW, CDKW, CEKT2, CEKT3}.

In previous articles we provided analytic results for the evaporation
of higher-dimensional black holes through the emission of scalars \cite{CEKT2},
fermions and gauge bosons \cite{CEKT3} on the brane. Here, we continue
our program of analytic study by focusing on
the emission of scalar degrees of freedom in the bulk from rotating
higher-dimensional black holes. In section 2 we set
up the general theoretical framework, considering the gravitational
background corresponding to a rotating higher-dimensional black
hole and writing down all relevant equations for a scalar field in
such a spacetime. In section 3 we provide an analytic solution of the
radial equation using a well-known matching technique.
This technique consists of, first, deriving the solution in the
{\textit{near horizon regime}}, then deriving the equivalent {\textit{far
field}} limit, before finally stretching and matching the two
forms in an intermediate zone. In this way an analytic expression for
the radial part of the field valid throughout the entire spacetime is
constructed. The solution obtained in this way is valid in the low-energy
and low-angular-momentum approximation. Our solution is used in section 4
to calculate  the absorption probability which characterises the emitted
Hawking radiation. We also perform a complimentary numerical analysis,
valid in the same regime, and compare the two sets of results. Plots are
depicted for various dimensional spacetimes and angular momentum modes.
In section 5 we compute the energy emission rate and the brane-to-bulk
ratio of the energy emission rates, and provide corresponding
plots. Finally, in section 6 we present our conclusions.

\section{General Framework}

The gravitational background around a $(4+n)$-dimensional, neutral, rotating
black hole is described by the well-known Myers-Perry solution \cite{MP}.
Here we will be interested in black holes created by an on-brane collision
of particles, and may therefore assume that the corresponding metric will
depend only on one non-zero angular momentum component, parallel to our brane.
Then, the line-element takes the form
\begin{eqnarray}
&~& \hspace*{-3cm}ds^2 = -\biggl(1-\frac{\mu}{\Sigma\,r^{n-1}}\biggr) dt^2 -
\frac{2 a \mu \sin^2\theta}{\Sigma\,r^{n-1}}\,dt \, d\varphi
+\frac{\Sigma}{\Delta}\,dr^2 +\Sigma\,d\theta^2 \nonumber \\[2mm]
\hspace*{2cm}
&+& \biggl(r^2+a^2+\frac{a^2 \mu \sin^2\theta}{\Sigma\,r^{n-1}}\biggr)
\sin^2\theta\,d\varphi^2 + r^2 \cos^2\theta\, d\Omega_{n}^2,
\label{rot-metric}
\end{eqnarray}
where
\begin{equation}
\Delta = r^2 + a^2 -\frac{\mu}{r^{n-1}}\,, \qquad
\Sigma=r^2 +a^2\,\cos^2\theta\,,
\label{Delta}
\end{equation}
and $d\Omega_n^2(\theta_1,\theta_2,\ldots,\theta_{n-1},\phi)$ is
the line-element on a unit $n$-sphere. The black hole's mass
$M_{BH}$ and angular momentum $J$ are related to the
parameters $a$ and $\mu$ as follows
\begin{equation}
M_{BH}=\frac{(n+2) A_{n+2}}{16 \pi G}\,\mu\,,  \qquad
J=\frac{2}{n+2}\,M_{BH}\,a\,, \label{def}
\end{equation}
with $G$ being the $(4+n)$-dimensional Newton's constant, and
$A_{n+2}$ the area of a $(n+2)$-dimensional unit sphere given by
$A_{n+2}=2 \pi^{(n+3)/2}/\Gamma[(n+3)/2]$.
Finally, the black hole's horizon radius $r_h$ follows from
the equation $\Delta(r_h)=0$, and is found to satisfy the
equation: $r_h^{n+1}=\mu/(1+a_*^2)$, where $a_*=a/r_h$.

In this work we will study the emission of Hawking radiation, in the form
of scalar fields, in the higher-dimensional space. We therefore need to
consider the equation of motion of a massless scalar field propagating in
the gravitational background (\ref{rot-metric}). This is given by
\beq
\frac{1}{\sqrt{-g}}\,\partial_\mu\left(\sqrt{-g}\,g^{\mu \nu}\partial_\nu \Phi\right)=0\,,
\eeq
where
\beq
\sqrt{-g}=\Sigma \sin\theta \, r^{n}\cos^{n}\theta \prod_{i=1}^{n-1} \sin^{i}\theta_i\,.
\eeq
The above equation can be separated by assuming the factorised ansatz
\beq
\Phi = e^{-i\omega t}e^{im\varphi}R(r)\,S(\theta)\,Y_{jn}(\theta_1,\ldots,\theta_{n-1},\phi) \,,
\eeq
where $Y_{jn}(\theta_1,\ldots,\theta_{n-1},\phi)$ are the hyperspherical harmonics on
the $n$-sphere that satisfy the equation \cite{Muller}
\begin{equation}
\sum_{k=1}^{n-1} \frac{1}{\prod_{i=1}^{n-1} \sin^i\theta_i}\,\partial_{\theta_k}
\left[\left(\prod_{i=1}^{n-1} \sin^i\theta_i\right)\,\frac{\partial_{\theta_k} Y_{jn}}
{\prod_{i>k}^{n-1} \sin^2\theta_i}\right] + \frac{\partial_{\phi\phi} Y_{jn}}
{\prod_{i=1}^{n-1} \sin^2\theta_i}+j(j+n-1)\,Y_{jn}=0\,.
\label{hyper-eq}
\end{equation}
The functions $R(r)$ and $S(\theta)$ are then found as the solutions to the
decoupled equations
\begin{equation}
\frac{1}{r^n}\,\partial_r \left(r^n\Delta\,\partial_r R\right) + \left(\frac{K^2}{\Delta} -
\frac{j(j+n-1)a^2}{r^2} - \Lambda_{j\ell m}\right)R = 0 \,,\label{eq:radial}
\end{equation}
\begin{equation}
\frac{1}{\sin\theta \cos^n \theta}\,\partial_\theta\left(\sin\theta \cos^n \theta
\partial_\theta S\right) +\left(\omega^2 a^2 \cos^2\theta - \frac{m^2}{\sin^2\theta}
- \frac{j(j+n-1)}{\cos^2\theta} + E_{j\ell m}\right)S = 0 \,, \label{eq:hdspheroidal}
\end{equation}
respectively, that first appeared in the literature in Ref. \cite{IMU}. In the
above,
\begin{equation}
K=(r^2+a^2)\,\omega-am\,,  \qquad
\Lambda_{j\ell m}=E_{j\ell m}+a^2\omega^2-2am\omega\,.
\end{equation}

The angular eigenvalue $E_{j\ell m} (a\omega)$ provides a link
between the angular and radial equation and, as in the case of
on-brane emission \cite{CEKT2, IOP, HK2, IOP2, DHKW}, there is
no closed analytic form for its value. It may however be expressed
as a power series in $a \omega$ \cite{Eigenvalues1,Eigenvalues2}.
For the purpose of our analysis, valid only in the
low-$\omega$ and low-$a$ limit, it suffices to keep a finite
number of terms and we will therefore truncate the series at the
5th order\footnote{The expression we use for $E_{j\ell
m}$ is based on the analysis of \cite{Eigenvalues1} but disagrees
slightly with the version given there as the sign of the second order
term is reversed. This is necessary so that, in the limit $j,n\rightarrow
0$, the expression for $E_{j\ell m}$ correctly reproduces the on-brane
eigenvalues that have appeared in the literature
previously.}. In order for the power series to converge in the limit
$a\omega\rightarrow0$ by terminating at finite order, a
number of restrictions are imposed on the allowed values of the
integer parameters $(j,\ell,m)$ specifying the emission mode\,: in
general, $m$ may take any integer value and $j$ and $\ell$ any
positive or zero integer value providing \cite{Eigenvalues1}
\begin{equation} \label{eq:restrictions}
\ell\geq j+|m| \qquad \textrm{and}  \qquad
\frac{\ell-(j+|m|)}{2} \in \{0,\mathbb{Z}^+\}\,.
\end{equation}

By using the power series form of the angular eigenvalues, we may
now proceed to solve Eq. (\ref{eq:radial}) analytically. The solution for
the radial function $R(r)$ will help us determine the absorption probability
$|{\cal A}_{j \ell m}|^2$ for the propagation of a massless scalar field in the bulk,
a quantity that characterises the Hawking radiation emission rates of the
black hole.

\section{Analytic Solution}

In order to obtain a solution to the radial equation (\ref{eq:radial}), we
will use a well-known approximate method: we will first solve the equation close
to the horizon of the black hole ($r\simeq r_h$), and then far away from it
($r \gg r_h$). Finally, we will smoothly match the two solutions
in the intermediate zone, thus creating an analytical solution
for the whole radial regime.

Starting from the near-horizon regime, we perform the
following transformation of the radial variable \cite{CEKT2,CEKT3}
\beq
r \rightarrow f(r) = \frac{\Delta(r)}{r^2+a^2} \,\,\Rightarrow\,
\frac{df}{dr}=(1-f)\,r\,\frac{A(r)}{r^2+a^2}\,,
\eeq
where $A(r) \equiv (n+1)+(n-1)\,a^2/r^2$. Then, near the horizon ($ r \simeq
r_h$), Eq. (\ref{eq:radial}) takes the form
\begin{equation}
f\,(1-f)\,\frac{d^2R}{df^2} + (1-D_*\,f)\,\frac{d R}{df}
+ \biggl[\,\frac{K^2_*}{A_*^2\,f (1-f)}-\frac{\left(j(j+n-1)a_*^2 +
\Lambda_{j\ell m}\right)\,(1+a_*^2)}{A_*^2\,(1-f)}\,\biggr] R=0\,,
\label{eq:NH-1}
\end{equation}
where $A_*$ and $K_*$ are given by
\begin{eqnarray}
A_* = (n+1) + (n-1)a_*^2, \qquad
K_* = (1+a_*^2) \omega_* - a_* m\,,
\end{eqnarray}
and
\beq
D_* \equiv 1 - \frac{4 a_*^2}{A_*^2}\,.
\label{eq:Dstar}
\eeq
By employing the transformation
\beq R_{NH}(f)=A_- f^{\alpha}\,(1-f)^\beta\,F(a,b,c;f)\,,
\eeq
equation (\ref{eq:NH-1}) can be brought to the form of a hypergeometric
differential equation \cite{Abramowitz}, with $a=\alpha + \beta +D_*-1$,
$b=\alpha + \beta$, and $c=1 + 2 \alpha$. In addition, the parameters
$\alpha$ and $\beta$ are given by
\begin{eqnarray}
&~& \hspace*{-1.cm} \alpha = \pm \frac{i K_*}{A_*}\,, \label{alpha}\\
&~& \hspace*{-1.cm} \beta  = \frac{1}{2}\,\biggl[\,(2-D_*) \pm
\sqrt{(D_*-2)^2 - 4\biggl[\frac{K_*^2 - (j(j+n-1)a_*^2 +
\Lambda_{j\ell m})\,(1+a_*^2)}{A_*^2}\biggr]}\,\,\biggr].
\label{beta}
\end{eqnarray}
The general solution of the radial equation (\ref{eq:NH-1}) may be written
in terms of the hypergeometric function $F$ as follows
\begin{eqnarray}
&& \hspace*{-1cm}R_{NH}(f)=A_-\,f^{\alpha}\,(1-f)^\beta\,F(a,b,c;f)
\nonumber \\[1mm] && \hspace*{2cm} +\,
A_+\,f^{-\alpha}\,(1-f)^\beta\,F(a-c+1,b-c+1,2-c;f)\,.
\label{NH-gen}
\end{eqnarray}
In this general near-horizon solution, we must now impose the
boundary condition that no outgoing modes exist near the black
hole's horizon. In the limit $r\rightarrow r_h$ we get $f(r)
\rightarrow 0$, and the near-horizon solution (\ref{NH-gen})
becomes
\beq R_{NH}(f) \simeq A_-\,f^{\pm i K_*/A_*} + A_+\,f^{\mp
i K_*/A_*}= A_-\,e^{\pm i ky} + A_+\,e^{\mp i ky }\,,
\end{equation}
where
\beq k  \equiv \omega - m \Omega =\omega - \frac{m a}{r_h^2 +a^2}\,,
\label{k} \eeq
and $y$ is a tortoise-like coordinate defined by
$y=r_h (1+a_*^2)\ln(f)/A_*$. By imposing this boundary condition, we
can set either $A_-=0$ or $A_+=0$, depending on the choice for $\alpha$.
The two choices are clearly equivalent, so we choose
$\alpha=\alpha_-$, which imposes $A_+=0$. This brings the near-horizon
solution to the final form
\beq R_{NH}(f)=A_-\,f^{\alpha}\,(1-f)^\beta\,F(a,b,c;f)\,.
\label{NH-final} \eeq
Finally, the convergence criterion for the hypergeometric function,
i.e. $Re(c-a-b)>0$, should also be applied, leading
us to choose $\beta=\beta_-$.


We now focus our attention on the far-field regime, $r \gg r_h$. In
this limit, the substitution $R(r) = r^{-\left(\frac{n+1}{2}\right)} \tilde R(r)$
brings Eq. (\ref{eq:radial}) into the form of a Bessel equation
\cite{Abramowitz}, in terms of $z=\omega r$,
\beq
\frac{d^2\tilde R}{dz^2}+ \frac{1}{z}\frac{d \tilde R}{dz}+\left(1 -
\frac{E_{j\ell m}+a^2\omega^2+ \left(\frac{n+1}{2}\right)^2}{z^2} \right)\tilde R=0 \,.
\label{FF-eq}
\eeq
Thus, the solution in the far-field regime may be written as
\beq
R_{FF}(r)=\frac{B_1}{r^{\frac{n+1}{2}}}\,J_{\nu}\,(\omega r) +
\frac{B_2}{r^{\frac{n+1}{2}}}\,Y_{\nu}\,(\omega r) \,,
\label{FF}
\eeq
with $J_\nu$ and $Y_\nu$ Bessel functions of the first and second kind,
respectively, and $\nu=\sqrt{E_{j\ell m}+a^2\omega^2+\left(\frac{n+1}{2}\right)^2}$.

In order to construct a full analytic solution valid for all $r$, we must
smoothly match the two asymptotic solutions (\ref{NH-final}) and (\ref{FF}) in an
intermediate regime. To this end, we first focus on the near-horizon solution
(\ref{NH-final}), and shift the argument of the hypergeometric function
from $f$ to $1-f$ by using the following relation \cite{Abramowitz}
\bea
\hspace*{-2mm}R_{NH}(f)&=&A_- f^\alpha\,(1-f)^\beta\,\Biggl[\,
\frac{\Gamma(c)\,\Gamma(c-a-b)}
{\Gamma(c-a)\,\Gamma(c-b)}\,F(a,b,a+b-c+1;1-f)  \\[1mm]
&+& (1-f)^{c-a-b}\,\frac{\Gamma(c)\,\Gamma(a+b-c)}
{\Gamma(a)\,\Gamma(b)}\,F(c-a,c-b,c-a-b+1;1-f)\,\Biggr]. \nonumber
\label{NH-shifted} \eea
The function $f(r)$ may be alternatively written as
\beq f(r)=1-\frac{\mu}{r^{n-1}}\,\frac{1}{r^2+a^2}=
1-\biggl(\frac{r_h}{r}\biggr)^{n-1}\,\frac{(1+a_*^2)}
{(r/r_h)^2+a_*^2}\,. \label{f-far} \eeq
In the limit $r \gg r_h$, the $(r/r_h)^2$ component in the denominator
of the second term dominates, and $f(r)$ goes to unity for
$n \geq 0$. Then, the near-horizon solution (\ref{NH-shifted})
can be written as
\beq R_{NH}(r) \simeq A_1\, r^{\,-(n+1)\,\beta} +
A_2\,r^{\,(n+1)\,(\beta + D_*-2)}\,, \label{NH-stretched} \eeq
with $A_1$ and $A_2$ defined as \bea A_1 &=& A_-
\left[(1+a_*^2)\,r_h^{n+1}\right]^\beta \,
\frac{\Gamma(c)\Gamma(c-a-b)}{\Gamma(c-a)\Gamma(c-b)}\,, \nonumber\\[1mm]
A_2 &=& A_- \left[(1+a_*^2)\,r_h^{n+1}\right]^{-(\beta + D_*-2)}
\, \frac{\Gamma(c)\Gamma(a+b-c)}{\Gamma(a)\Gamma(b)}\,. \eea

Next, we expand the far-field solution to small values of $r$, by
taking the limit $r\rightarrow 0$ in Eq. (\ref{FF}) :
\beq
R_{FF}(r) \simeq \frac{B_1\left(\frac{\omega r}{2}
\right)^\nu}{r^{\frac{n+1}{2}} \,\Gamma(\nu+1)}-
\frac{B_2}{\pi \, r^{\frac{n+1}{2}}}\,\frac{\Gamma(\nu)}{\left(\frac{\omega r}{2}
\right)^{\nu}} \,.
\label{FF-stretched}
\eeq
Then, taking the small $a_*$ and $\omega_*$ limit in the power coefficients
of $r$ -- so that we can ignore terms of order $(\omega_*^2, a_*^2, a_*\omega_*)$
or higher -- we can achieve exact matching since
\bea
-(n+1)\,\beta &\simeq& \ell + {\cal O}(\omega_*^2, a_*^2, a_*\omega_*)\,, \nonumber \\[2mm]
(n+1)\,(\beta+D_*-2) &\simeq& -(\ell+n+1) + {\cal O}(\omega_*^2,
a_*^2, a_*\omega_*)\,,
\nonumber \\[1mm] \nu &\simeq& \ell+\frac{n+1}{2} + {\cal O}(a_*^2\omega_*^2)\,.
\eea
We would like to stress here that, in order to achieve a higher level of accuracy
in our analysis, no expansion is performed in the arguments of the gamma functions,
and terms to order $(a\omega)^4$ are retained in the expansion of the eigenvalues.
Then, the matching of the two asymptotic solutions leads to the constraint
\bea &~& \hspace*{-1.5cm}B \equiv \frac{B_1}{B_2} = -\frac{1}{\pi}
\left(\frac{2} {\omega
r_h\,(1+a_*^2)^\frac{1}{n+1}}\right)^{2\ell+n+1}
\sqrt{E_{j\ell m}+a^2\omega^2+\left(\frac{n+1}{2}\right)^2} \nonumber \\[2mm]
&& \hspace*{-0.6cm} \times\,\frac{\Gamma^2\left(\sqrt{E_{j\ell m}+a^2\omega^2+
\left(\frac{n+1}{2}\right)^2}\right) \Gamma(\alpha+\beta + D_* -1)\,
\Gamma(\alpha+\beta)\, \Gamma(2-2\beta - D_*)}{\Gamma(2\beta + D_*-2)\,
\Gamma(2+\alpha -\beta - D_*)\,\Gamma(1+\alpha-\beta)} \,. \label{eq:Beq}
\eea
The above guarantees the existence of a smooth, analytic solution for the radial
part of the wavefunction for all $r$, valid in the low-energy and low-rotation limit.


\section{The Absorption Probability}

With the solution to the radial equation (\ref{eq:radial}) at our disposal,
we may now compute the absorption probability. To this end, we expand the
far-field solution (\ref{FF-eq}) for $r\rightarrow \infty$. By
using standard formulae for the Bessel functions \cite{Abramowitz}, we find
\bea R_{FF}(r) &\simeq&
\frac{1}{r^\frac{n+2}{2}\sqrt{2\pi\omega}}\left[(B_1+iB_2)\,
e^{-i\,\left(\omega r - \frac{\pi}{2}\,\nu - \frac{\pi}{4}\right)}
+ (B_1-iB_2)\,e^{i\,\left(\omega r - \frac{\pi}{2}\,\nu -
\frac{\pi}{4}\right)} \right]\nonumber \\
&=& A_{in}^{(\infty)}\,\frac{e^{-i\omega r}}{r^\frac{n+2}{2}} +
A_{out}^{(\infty)}\,\frac{e^{i\omega r}}{r^\frac{n+2}{2}} \,. \eea
We see that for large distances from the black hole, the solution
reduces to an incoming and an outgoing spherical wave, allowing us
to compute the absorption probability from the ratio of their
amplitudes:
\beq \left|{\cal A}_{j \ell m}\right|^2 =
1-\left|\frac{A_{out}^{(\infty)}}{A_{in}^{(\infty)}}\right|^2
= 1-\left|\frac{B_1-iB_2}{B_1+iB_2}\right|^2 =
\frac{2i\left(B^*-B\right)}{B B^* + i\left(B^*-B\right)+1}\,.
\label{Absorption} \eeq
The above equation, in conjunction with Eq. (\ref{eq:Beq}), is our main analytic
result for the absorption probability characterising the emission of massless
scalar fields in the bulk, from a rotating, uncharged black hole, in the low-energy
and low-angular momentum limit.

\begin{figure}[t]
  \begin{center}
  \mbox{\includegraphics[width = 0.7 \textwidth] {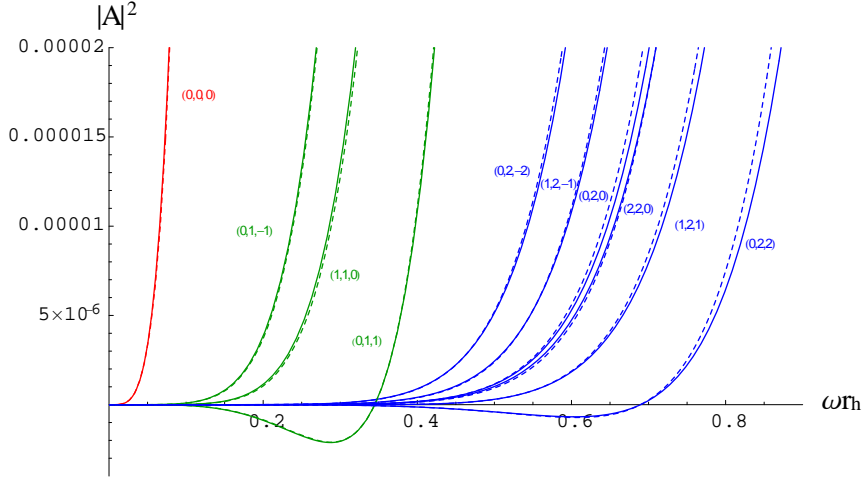}}
\hspace*{-0.1cm}
    \caption{ Absorption probabilities $\left|{\cal A}_{j \ell m}\right|^2$ for a bulk
    scalar field, for $n=2$ , $a_*=0.4$, and various combinations of $(j,\ell,m)$.}
   \label{jlmcomp}
  \end{center}
\end{figure}

In Fig. (\ref{jlmcomp}) we plot the absorption probability for the first
partial waves for $n=2$, $a_*=0.4$, with the values of $j,\ell,m$ obeying
the restrictions (\ref{eq:restrictions}). One may easily observe the dominance
of the first partial wave $j=\ell=m=0$ over all others, and the suppression
of $\left|{\cal A}_{j \ell m}\right|^2$ as the values of the angular momentum
numbers increase. In the plot, we also see the appearance of superradiance
\cite{super} for modes with positive $m$, where the absorption probability
takes negative values. Figure (\ref{jlmcomp}) actually depicts two sets of
curves: the first, denoted by solid lines, gives the value of
$\left|{\cal A}_{j \ell m}\right|^2$ that follows from our analytic result
(\ref{Absorption}), while the second set, denoted by dashed lines, gives
the value following from integrating the radial equation (\ref{eq:radial})
numerically\footnote{In the numerical integration of Eq. (\ref{eq:radial})
the power series expansion of the eigenvalue $E_{j\ell m}$ \cite{Eigenvalues1}
up to 5th order, valid in the low-$a\omega$ limit, was again used.}.
As in the case of scalar
\cite{CEKT2} and higher-spin fields \cite{CEKT3} propagating on the brane,
our approximate analytic method leads to results that are
in excellent agreement with the exact numerical ones, not only in the
low-energy regime but beyond this also.

\begin{figure}[t]
  \begin{center}
  \mbox{\includegraphics[width = 0.5 \textwidth] {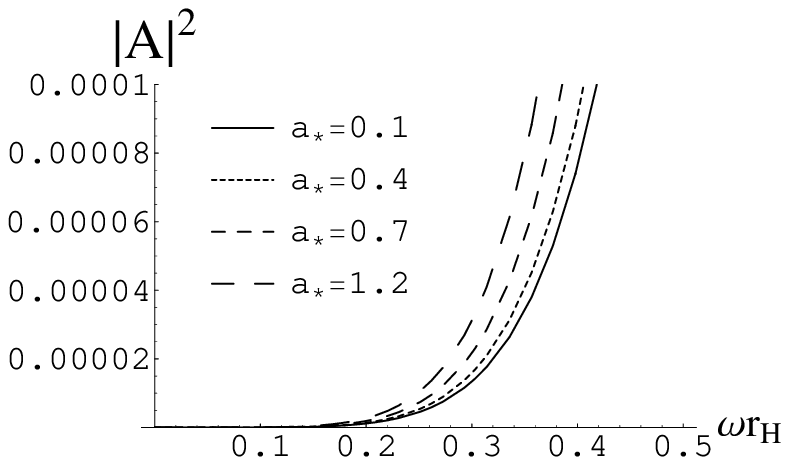}}
\hspace*{-0.3cm} {\includegraphics[width = 0.5 \textwidth]
{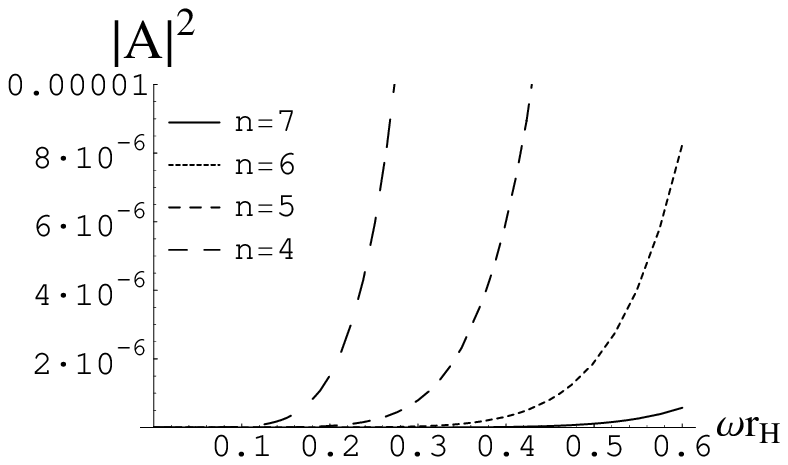}}
    \caption{ Absorption probabilities for the bulk scalar mode $j=\ell=m=0$, for
\textbf{(a)} $n=5$ and various $a_*$, and \textbf{(b)} $a_*=0.5$ and various $n$.}
   \label{ancomp}
  \end{center}
\end{figure}

Focusing on the dominant first partial wave, in Fig. \ref{ancomp}(a) we
demonstrate the dependence of the absorption probability on the rotation
parameter $a_*$, for fixed $n=5$. One may clearly see that an increase in
the rotation of the black hole causes an enhancement in the value of
$\left|{\cal A}_{j\ell m}\right|^2$ in the low- and intermediate-energy regimes.
In Fig. \ref{ancomp}(b), we investigate instead the effect of the number of
extra dimensions on the value of the absorption probability, while we fix
$a_*=0.5$. As is clear from the plot, the value of $\left|{\cal A}_{j \ell m}\right|^2$
in the low-energy regime is strongly suppressed as $n$ increases.
The same behaviour in terms of $n$ was found for bulk scalar fields
propagating in the background of a higher-dimensional, spherically-symmetric
black hole \cite{HK1}.

An interesting question is how the absorption probabilities for brane and
bulk scalar fields in a rotating black-hole background compare.
By examining the results presented in this section and
in \cite{CEKT2}, one may easily conclude that the absorption probabilities
for both species of scalar fields are enhanced as the black hole rotation
parameter increases. In contrast to this, the value of the absorption
probability increases with $n$ for brane scalars, while
it decreases for bulk scalar fields. Important conclusions can
also be drawn by directly comparing Fig. \ref{jlmcomp} with the corresponding figure
in \cite{CEKT2}: for the same values of $n$ and $a_*$,
the absorption probability for brane scalar fields is consistently
larger than that for bulk scalars by almost 3 orders of magnitude,
both for superradiant and non-superradiant modes. The same observation
was made in \cite{Jung-rot} in the 5-dimensional case -- here, we
have shown that this behaviour persists for higher values of $n$ also.

A compact analytic expression for the absorption probability may be derived
in the very low-energy limit. For $\omega \rightarrow 0$, we obtain
$B \propto 1/\omega^{2\ell+n+1}$, therefore
\beq \left|{\cal A}_{j\ell m}\right|^2 \simeq  2i\left(\frac{1}{B}
- \frac{1}{B^*}\right)\,. \eeq
Substituting for $B$ from Eq. (\ref{eq:Beq}), we take
\bea \left|{\cal A}_{j\ell m}\right|^2 &=&
\frac{-2i\pi\,\left(\omega r_h/2\right)^{2\ell+n+1}}
{(\ell+\frac{n+1}{2})\,\Gamma^2(\ell+\frac{n+1}{2})} \frac{\Gamma(2\beta
+D_*-2)}{(1+a_*^2)^{-\frac{2\ell+n+1}{n+1}}\,\Gamma(2-2\beta -D_*)} \times\nonumber\\[1mm]
&&\left[\frac{\Gamma(2+\alpha-\beta-D_*)\,\Gamma(1+\alpha-\beta)}{\Gamma(\alpha+\beta+D_*-1)\,
\Gamma(\alpha+\beta)}-\frac{\Gamma(2-\alpha-\beta-D_*)\,\Gamma(1-\alpha-\beta)}
{\Gamma(-\alpha+\beta+D_*-1)\,\Gamma(-\alpha+\beta)}\right]\,.
\nonumber \eea
Focusing our attention on the dominant mode $j=\ell=m=0$, and performing
an analysis similar to that in \cite{CEKT2}, where the absorption probability
of the dominant scalar mode on the brane was found in the same limit, we obtain
the result
\beq
\left|{\cal A}_{0}\right|^2 = \frac{4\pi (1+a_*^2)^2(\omega r_h)^{n+2}}
{A_* \, 2^n(n+1)\Gamma^2\left(\frac{n+1}{2}\right)(2-D_*)} + \ldots \,\,.
\eeq

\begin{figure}[t]
  \begin{center}
  \includegraphics[width = 0.7 \textwidth] {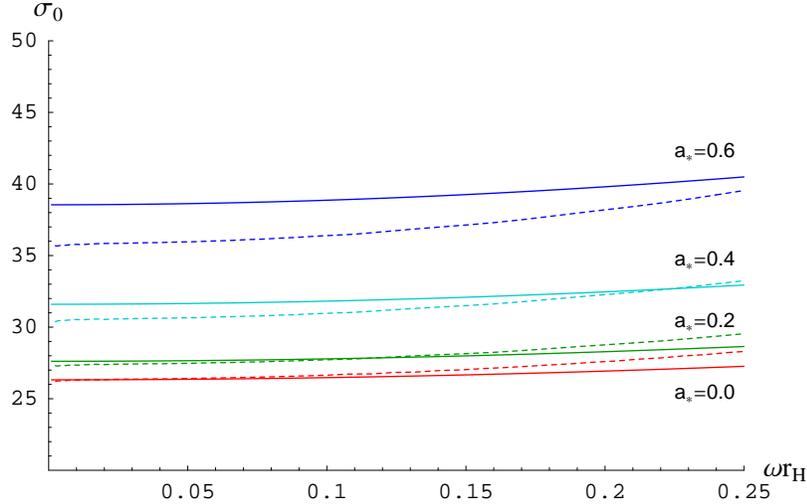}
    \caption{ Absorption cross-section for the bulk scalar mode $j=\ell=m=0$, for $n=2$
    and various values of $a_*$. }
   \label{sigma}
  \end{center}
\end{figure}

The above result allows us also to compute the absorption cross-section $\sigma_0$
for the dominant scalar bulk mode in the asymptotic low-energy regime. By adapting the
formula of Ref. \cite{cross-section} to the geometrical set-up of our analysis,
we may write
\beq
\sigma_{j\ell m}(\omega) = \frac{2^n}{\pi} \Gamma^2\left(\frac{n+3}{2}\right)
\frac{A_H}{(\omega r_h)^{n+2}}\,\frac{N_j}{(1+a_*^2)}\left|{\cal A}_{j\ell m}\right|^2 \,,
\label{cross-section}
\eeq
where now
\beq
N_j = \frac{(2j+n-1)(j+n-2)!}{j!\,(n-1)!}\,, \qquad
A_H = \frac{ 2\pi^{\frac{n+3}{2}}r_h^n\,(r_h^2+a^2)}{\Gamma\left(\frac{n+3}{2}\right)}
\label{Nj}
\eeq
are the multiplicity of the $j$th partial wave in the expansion of the wave function over
the hyperspherical harmonics on the $n$-sphere, and the horizon area of the $(4+n)$-dimensional
rotating black hole, respectively. Substituting for the absorption coefficient, we obtain
\beq
\sigma_{0}(\omega) \simeq  \frac{(n+1)(1+a_*^2) A_H}{A_*(2-D_*)} + \ldots\,.
\eeq
For $a_* \rightarrow 0$, the above reduces to $A_H$, a behaviour that was found
in \cite{HK1}. For $a_* \neq 0$, the numerical results (dashed lines) presented in
Fig. \ref{sigma} confirm that, also here, the low-energy limit of the cross-section
tends to the area of the corresponding rotating black hole. The solid lines demonstrate
the agreement of our analytic results with the numerical ones for small values of
$a_*$, and the expected deviation for large values of the rotation parameter.


\section{Energy Emission Rate}

Having found the absorption probability (Eq. \ref{Absorption}), we
can now proceed to compute the rate of energy emission of massless scalar fields
in the bulk. This is given by the expression
\beq \frac{d^2 E}{dt d\omega} = \frac{1}{2\pi}\sum_{j, \ell,
m}\frac{\omega}{\exp\left[k/T_\text{H}\right]-1}\,N_j
\left|{\cal A}_{j\ell m}\right|^2 \,. \label{dedt}\eeq
The above differs from the 4-dimensional \cite{OW} expression in the presence
of an additional sum over the new angular momentum number $j$, and from
the 5-dimensional \cite{FS-rot} one in the introduction of the
multiplicity of states $N_j$ (\ref{Nj}) following from the expansion of
the wavefunction of the field in the $n$-dimensional sphere. The parameter
$k$ is defined in Eq. (\ref{k}), while the temperature of the higher-dimensional,
rotating black hole is
\begin{equation}
T_\text{H}=\frac{(n+1)+(n-1)a_*^2}{4\pi(1+a_*^2)r_{h}}\,.
\end{equation}

A useful check that may convince us of the validity of the above emission
rate is to take the non-rotating limit. Then, Eq. (\ref{dedt}) for the energy rate
should reduce to the well-known result for bulk scalar emission from a
$(4+n)$-dimensional Schwarzschild black hole \cite{KMR1, HK1}
\beq \frac{d^2 E}{dt d\omega} = \frac{1}{2\pi}\sum_{\ell}\frac{\omega}
{\exp\left[\omega/T_\text{H}\right]-1}\,N_\ell
\left|{\cal A}_{\ell}\right|^2 \,, \label{dedtS}\eeq
where now $N_\ell$ is the degeneracy of the $\ell$th mode of the harmonics on the
$(n+2)$-sphere
\beq N_\ell =\frac{(2\ell+n+1)(\ell+n)!}{\ell!\,(n+1)!}\,.
\eeq
In the limit $a \rightarrow 0$, we get $k=\omega$, and the absorption
probability becomes independent of both $m$ and $j$, retaining a dependence
only on the principal quantum number $\ell$. Then, Eq. (\ref{dedt})
matches Eq. (\ref{dedtS}) providing the following relation holds
\beq
\sum_{j,m} N_j=\sum\limits_{j = 0}^\ell \,(\ell - j + 1)\,\frac{(2j + n - 1)(j + n - 2)!}
{j!\,(n - 1)!} \equiv N_\ell\,. \label{1}
\eeq
In the second part of the above equation, we have used the fact that,
according to the restrictions (\ref{eq:restrictions}) imposed on the quantum
numbers, for each value of $(j,\ell)$, $m$ may take $\ell-j+1$ values,
and for each value of $\ell$, $j$ may take the values $0\leq j \leq \ell$.
In order to prove Eq. (\ref{1}), we rewrite the factor $2j+n-1$ as $(j+n-1)+ j$,
and split the sum in two parts. Then, if we further replace the index
$j$ by $i-n+1$ in the first sum and by $i-n+2$ in the second, the
middle part of Eq. (\ref{1}) takes the form
\beq
\sum\limits_{i = n -
1}^{\ell + n - 1} {(\ell + n - 1 - i + 1)\,C(i,n - 1)} +
\sum\limits_{i = n - 1}^{\ell + n - 2} {(\ell + n - 2 - i + 1)\,C(i,n - 1)}\,,
\label{inter}
\eeq
where $C(s,r)$ is the combination function, $C(s,r) = \frac{s!}{r!(s - r)!}$.
By using the identity
\beq \sum\limits_{i = r}^s {(s - i + 1)\,C(i,r) = C(s + 2,r+ 2)}\,,
\label{2} \eeq
the first sum in Eq. (\ref{inter}) reduces to $C(\ell+n+1,n+1)$, and the second
to $C(\ell + n,n + 1)$ - their sum can be easily shown to equal $N_\ell$.

\begin{figure}[t]
  \begin{center}
  \mbox{\includegraphics[width = 0.5 \textwidth] {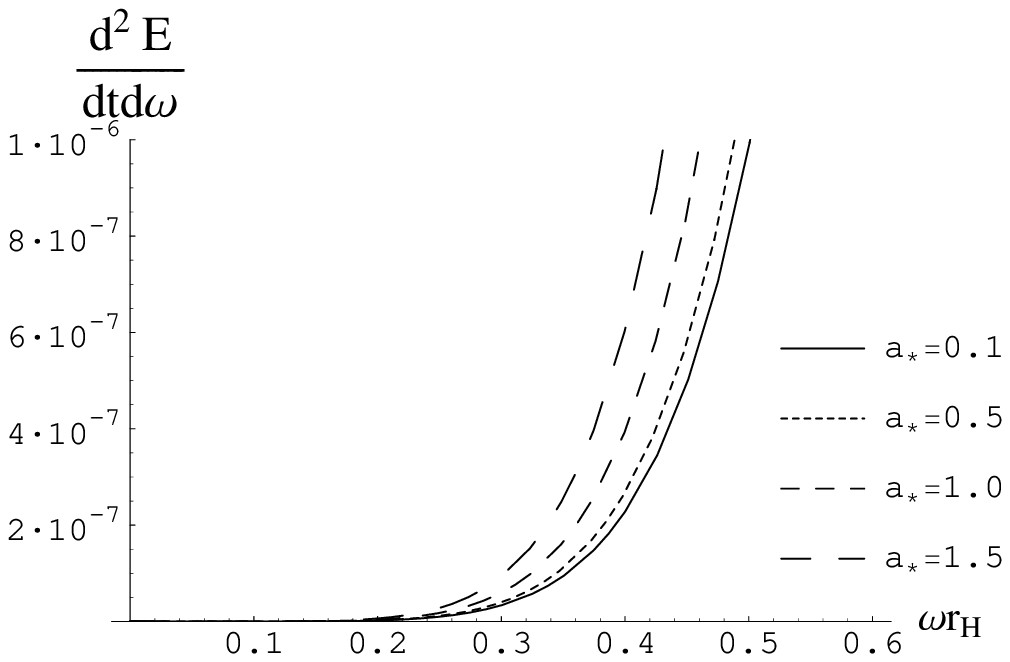}}
\hspace*{-0.3cm} {\includegraphics[width = 0.5 \textwidth]
{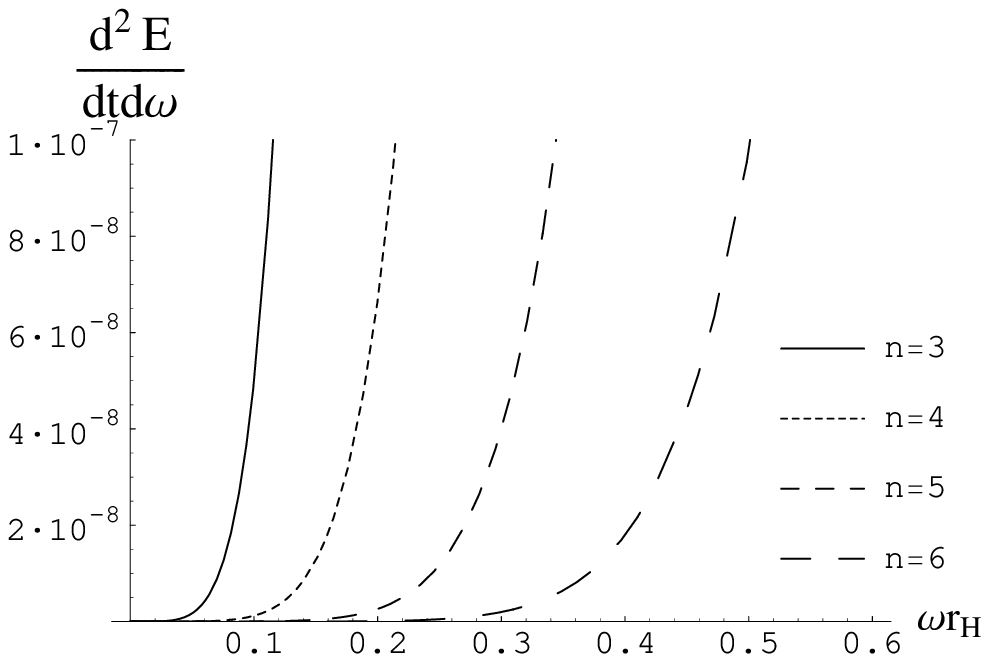}}
    \caption{ Energy emission rates for bulk scalar fields \textbf{(a)}
for $n=5$ and various $a_*$, and \textbf{(b)} for $a=0.5$ and various $n$.}
   \label{dedtvar}
  \end{center}
\end{figure}

By using Eq. (\ref{dedt}), in Figs. \ref{dedtvar}(a,b), we plot
the energy emission rate for a higher-dimensional rotating black hole in
the form of bulk scalar fields, as a function of the energy parameter
$\omega r_h$, and in terms of the angular momentum parameter
and number of extra dimensions, respectively. The profile exhibited by
the absorption probability is also observed here: the emission rate is
enhanced with $a_*$ in the low-energy regime, in analogy with results
for brane scalars, fermions and gauge bosons \cite{HK2,IOP2,DHKW,CKW,
CDKW, CEKT2,CEKT3}, while is suppressed in terms of $n$.
Drawing experience from previous studies, we expect the enhancement in
terms of $a_*$ to persist over the whole energy regime -- on the other
hand, it is quite likely, given the similarity of the results with those
for bulk scalar fields in a non-rotating background \cite{HK1}, that the
low-energy suppression with $n$ will be replaced by a strong enhancement
at the high-energy regime.

\begin{figure}[t]
  \begin{center}
  \mbox{\includegraphics[width = 0.5 \textwidth] {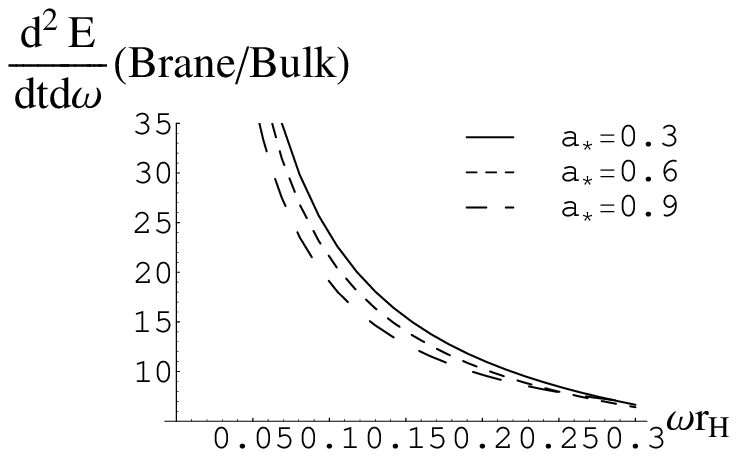}}
\hspace*{-0.3cm} {\includegraphics[width = 0.5 \textwidth]
{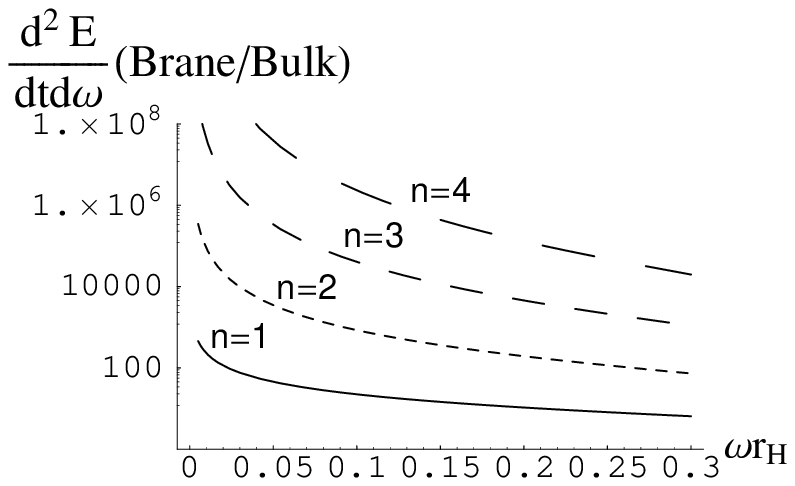}}
    \caption{ Brane-to-bulk ratio of the energy emission rates for scalar fields
   \textbf{(a)} for $n=1$ and various $a_*$, and \textbf{(b)} for $a=0.5$ and various $n$}
   \label{brane-to-bulk}
  \end{center}
\end{figure}

We should finally address the question of the brane-to-bulk ratio
of the rate of scalar field energy emission from a higher-dimensional rotating black hole. The
answer to this question is important as it will define the amount
of energy spent by the black hole in the observable brane
channel. Whereas the energy emission rate for bulk scalars is given
by Eq. (\ref{dedt}), the corresponding one for brane scalar degrees
of freedom is \cite{IOP, HK2, CDKW, IOP2}
\beq \frac{d^2 E}{dt d\omega} = \frac{1}{2\pi}\sum_{\ell,
m}\frac{\omega}{\exp\left[k/T_\text{H}\right]-1}\,
\left|{\cal A}_{\ell m}\right|^2 \,. \label{dedt-brane}\eeq
The comparison between the bulk and brane absorption
probabilities discussed in the previous section has given a
clear signal as to which emission is dominant, however the final
comparison should involve the total emission rates where the
different multiplicities of states have been taken into account.
By using Eqs. (\ref{dedt}) and (\ref{dedt-brane}), the brane-to-bulk
ratio for scalar fields from a rotating black hole is depicted
in Figs. \ref{brane-to-bulk}(a,b) in terms of the parameters $a_*$
and $n$. We may easily observe that although the exact value is both
$a_*$ and $n$-dependent, the ratio of the brane to bulk
emission rates always remains above unity, rendering the brane channel dominant.
We remind the reader that a similar conclusion was also
drawn in the case of a Schwarzschild-like higher-dimensional
black hole \cite{HK1}.

\section{Conclusions}

In this work we have performed an analytic study of the emission of scalar
fields in the bulk from a higher-dimensional rotating black hole. By solving
analytically the radial part of the equation of the scalar field, we have
constructed a smooth solution, valid in the low-energy and low-angular-momentum
regime. We were then able to compute the corresponding absorption probability,
and to examine its behaviour in terms of the particle's angular momentum numbers
and spacetime properties. The lowest scalar mode was found to be dominant,
with its absorption cross-section equal to the horizon of the
higher-dimensional rotating black hole in the extreme low-energy limit.
We demonstrated that the absorption probability for bulk scalar fields
is enhanced with increasing angular momentum of the black hole and suppressed
by the number of extra dimensions. We also performed a numerical analysis,
valid again in the low-$\omega$ and low-$a_*$ regime, and showed that our analytic
results are in excellent agreement with the exact numerical ones even up to
intermediate-energy regimes.

We then computed the energy emission rate in the bulk, and showed that
it exhibits the same behaviour as the absorption probability in terms
of $a_*$ and $n$. Finally, we calculated the brane-to-bulk ratio of energy
emission rates for scalar fields in a rotating, higher-dimensional background
and found that, in the low-$\omega$ and low-$a_*$ regime, it is always
larger than unity. This is due to the fact that the larger multiplicity of
scalar states in the bulk cannot counteract the fact
that the bulk absorption probability is approximately three orders of magnitude
smaller than the corresponding brane value in the low energy limit. It is
this dominance of the brane absorption probability that largely determines
the preference of the black hole to emit scalar field energy on the brane.
The complete radiation spectrum -- that may follow only from an exact,
numerical analysis involving both the angular and radial parts of the field
equation -- is necessary to decide whether the brane-to-bulk ratio remains
above unity over the entire frequency range.

\bigskip

{\bf Acknowledgments.} P.K. is grateful to M. Casals, S. Dolan and
E. Winstanley for use\-ful discussions. S.C and O.E. acknowledge
PPARC and I.K.Y. fellowships, respective\-ly. P.K. is funded by
the UK PPARC Research Grant PPA/A/S/2002/00350. K.T. and P.K.
acknowledge participation in the RTN networks
UNIVERSENET-MRTN-CT-2006035863-1 and MRTN-CT-2004-503369.

\end{document}